\documentclass[letterpaper, 10 pt, conference]{ieeeconf}  

\IEEEoverridecommandlockouts                              

\usepackage{graphics} 
\usepackage{epsfig} 
\usepackage{mathptmx} 
\usepackage{times} 
\usepackage{amsmath} 
\usepackage{amssymb}  
\usepackage{cite}
\usepackage{balance}

\title{\LARGE \bf
PID-GM: PID Control with Gain Mapping*
}

\author{Bo Zhu$^{1}$, Wei Yu$^{2}$ and Hugh H.T. Liu$^{3}$  
	\thanks{*This work was supported by the National Natural Science
		Foundation of China under Grant 61773095 and Grant 61973327}
	\thanks{$^{1}$ Bo Zhu is with the Center for Advanced Control and Smart Operations (CACSO), Nanjing University, Suzhou, 215163, China.
      {\tt\small zhubo@nju.edu.cn}}
	\thanks{$^{2}$  Wei Yu is with the Faculty of Data Science, City University of Macau, Macau SAR China, 999078, China.
      {\tt\small weiyu@cityu.edu.mo}}
	\thanks{$^{3}$ Hugh H.T. Liu is with Flight Systems and Control (FSC) lab, UTIAS, Toronto, Canada. 
		{\tt\small liu@utias.utoronto.ca}}%
}

\begin{document}

\maketitle
\thispagestyle{empty}
\pagestyle{empty}

\begin{abstract}

Proportional-Integral-Differential (PID) control is widely used in industrial control systems. However, up to now there are at least two open problems related with PID control. One is to have a comprehensive understanding of its robustness with respect to model uncertainties and disturbances. The other is to build intuitive, explicit and mathematically provable guidelines for PID gain tuning. In this paper, we introduce a simple nonlinear mapping to determine PID gains from three auxiliary parameters. By the mapping, PID control is shown to be equivalent to a new PD control (serving as a nominal control) plus an uncertainty and disturbance compensator (to recover the nominal performance). Then PID control can be understood, designed and tuned in a Two-Degree-of-Freedom (2-DoF) control framework. We discuss some basic properties of the mapping, including the existence, uniqueness and invertibility. Taking as an example the PID control applied to a general uncertain second-order plant, we prove by the singular perturbation theory that the closed-loop steady-state and transient performance depends explicitly on one auxiliary parameter which can be viewed as the virtual singular perturbation parameter (SPP) of PID control. All the three PID gains are monotonically decreasing functions of the SPP, indicating that the smaller the SPP is, the higher the PID gains are, and the better the robustness of PID control is. Simulation and experimental examples are provided to demonstrate the properties of the mapping as well as the effectiveness of the mapping based PID gain turning.

\end{abstract}

\section{INTRODUCTION}

Proportional-Integral-Differential (PID) controls are widely used in industrial control systems because of their simplicity and effectiveness. However, for the PID control applied to uncertain and disturbed plants, there are two fundamental issues that have not been well answered. One is to characterize the role of the involved three control components in guaranteeing uncertainty rejection performance or to improve its robustness in a systematic method \cite{skoczowski2005method} and \cite{alfaro2009robust}. The other is to seek an intuitive method to tune PID control gains so that a given performance specification is easily fulfilled without depending heavily on the trial and error fashion.

Some classic results on the disturbance rejection performance of integral control (IC) have been reported in \cite{goodwin2001}, \cite{aastrom2006advanced} and \cite{visioli2006}. 
Applications of IC to nonlinear systems can be found in
integrator backstepping \cite{krstic1995} and IC via linearization \cite{Khalil2002}. In many applications, a zero-error asymptotic tracking cannot be achieved due to model uncertainties or disturbances. Hence, one often turns to expect that the tracking error is ultimately bounded by a reasonably small bound \cite{Khalil2002}. However, it still lacks of a straightforward method for PID gain tuning to achieve a specified ultimate bound of control error. The challenge stems from the fact that each of PID gains may affect the ultimate bounds of control error, and it is difficult to describe the relationship between the gains and the ultimate bounds of control error in a simple and intuitive form.  Moreover, some understanding about the roles of P, I and D gains are conditionally effective and not always true. For instance, the steady-state errors can be reduced by setting a higher integral gain. However, this is not always the case because a too large integral gain could results in a more obvious oscillating phenomenon and large steady-state error.
 
Many works have been reported on the design of
uncertainty and disturbance estimators (UDEs) for robustness improvement in uncertain systems \cite{Zhong04}-\cite{Wen-Hua2012}. Recently, this robust control idea is applied to distributed multi-vehicle systems \cite{zhu2015CEP} and \cite{zhu2016robust}, and extended to the output feedback cases in recent work \cite{zhu2018IJRNC} and \cite{zhang2019TIM}. Both of IC and UDE-based control are appealing to engineers owing to the three main advantages over some modern approaches: 1) having a simple structure for easy understanding and implementation (the frequency-domain analysis and design are usually available), 2) avoiding the introduction of additional nonlinear factors that renders the system analysis complicated, and 3) producing continuous or smooth control signals. 

This paper aims to provide a simple gain mapping for PID controller, and mathematically prove that PID controller is equivalent to the combination of a new PD controller and a continuous UDE. This equivalence has been suggested in our earlier work \cite{Bohr15}, but a detailed theoretical analysis has not been provided and the considered uncertainties are limited to bounded exogenous disturbances in \cite{Bohr15}. In this paper, the uncertainty estimation performance is systematically analyzed by the well-known singular perturbation theory \cite{Khalil2002}. Particularly, generic uncertainties are considered, which are state-dependent and input-dependent signals. By the proposed mapping, the tuning process of the three PID gains is converted into a single-gain tuning for a fast estimator dynamics. Further, a linear
relationship between the single parameter and the ultimate bound of tracking error is built, indicating that the single-parameter tuning is straightforward and effective. The method is verified via both numerical
simulations and experimental results on the attitude trajectory tracking of a 3-DOF helicopter.


%


The major notations and symbols used in this paper are introduced as follows. For a vector $X\in R^{n\times 1}$ and a matrix $M=\left[ m_{ij}%
\right] \in R^{m\times n}$, ${{\left\Vert X\right\Vert }_{\infty} = \max\limits_{1\leq i\leq n}\left\vert {x_{i}}\right\vert }$, ${{\left\Vert
		X\right\Vert }_{2} = {\left( {\sum_{i=1}^{n}{\left\vert x_{i}\right\vert }^{2}}%
		\right) }^{\frac{1}{2}}}$, ${\left\Vert M\right\Vert }_{\infty
} = \max\limits_{1\leq i\leq m}\sum_{j=1}^{n}{\left\vert m_{ij}\right\vert }$,
${\left\Vert M\right\Vert }_{2} = \left( {\lambda _{max}\left( M^{T}M\right) }%
\right) ^{\frac{1}{2}}$. For $M \in R^{m\times m}$, $\lambda
_{max}(M)=\max\limits_{1\leq i\leq m}\left\vert \lambda _{i}\right\vert $
and $\lambda _{min}(M)=\min\limits_{1\leq i\leq m}\left\vert \lambda
_{i}\right\vert $, with $\lambda _{i}$ the eigenvalues of $M$. $I_{n}$
denotes the $n \times n$ identity matrix. $\mathbf{0}_{n}$ is the $n$%
-dimensional column vector with all elements $0$. $\left\vert a\right\vert $
denotes the absolute value (modulus) of real number $a$. For a scale time
function $u(t)$, $\left\Vert u\right\Vert _{\infty} \triangleq \underset{%
	t\geq 0}{\sup }\left\vert u\left( t\right) \right\vert $ and a bounded $u(t)$
is also denoted by $u\in \mathcal{L}_{\infty }$. A continuous function $%
\alpha: [0, a) \mapsto [0, \infty)$ is said to belong to class $\mathcal{K}$
if it is strictly increasing and $\alpha(0) = 0$ (Definitions 4.2 in \cite%
{Khalil2002}). A continuous function $\beta: [0, a) \times [0, a) \mapsto
[0, \infty)$ is said to belong to class $\mathcal{KI}$ if, for each fixed $s$%
, the mapping $\beta(r, s) $ belongs to class $\mathcal{K}$ with respect to $%
r$ and, for each fixed $r$, the mapping $\beta(r, s)$ is decreasing with
respect to $s$ and $\beta(r, s)\rightarrow 0$ as $s \rightarrow \infty$
(Definitions 4.3 in \cite{Khalil2002}).

Definition of $O\left(\epsilon\right)$: 
A vector function $f(t, \epsilon) \in R^n$ is said to be $O\left(\epsilon\right)$ over an interval $\left[t_1, t_2\right]$ if there exist positive constants $k$ and $\epsilon^{*}$ such that
\begin{equation}
\lVert f(t, \epsilon) \rVert \leq k\epsilon ~~~~\forall \epsilon \in \left[0, \epsilon^{*}\right], \forall t \in \left[t_1, t_2\right], \label{FirstAppro}
\end{equation}
where $\lVert \cdot \rVert$ is the Euclidean norm.

\section{PROBLEM FORMULATION}
We first show the difficulty in tuning PID gains for stability and steady-state performance specifications. Consider PID control for the following uncertain second-order plant: 
\begin{equation}
\ddot{q}\left( t\right) = u\left( t\right) +%
d\left(q, \dot{q}, u, t\right),
\label{Plant_3}
\end{equation}%
where $\left(q, \dot{q}\right) ^{T}\in R$ is the
state, $q$ is the controlled output, $u\in
{R}$ is the control input, $d\left(
q, \dot{q}, u, t\right) $ is the unknown lumped uncertainty and disturbance (LUD). Suppose LUD has the following general linear form:
\begin{equation}
d\left(q, \dot{q}, u, t\right) = a_{1}%
q\left(t\right)+ a_{2}%
\dot{q}\left(t\right)
+ b u\left( t\right) + w\left( t\right),
\label{Plant_4}
\end{equation}%
where $a_{1}$, $a_{2}$ and $b$ are unknown real coefficients, and $w(t)$ denotes the unknown time-varying exogenous disturbance. 
From (\ref{Plant_4}), the initial value of LUD, denoted by $d_0$, is
\begin{eqnarray}
d_0 &\triangleq& d\left(q\left(0\right),\dot{q}\left(0\right), u\left( 0\right), 0\right) \notag \\
&=& a_{1}%
q\left(0\right) + a_{2}\dot{q}\left(0\right)
+ b u\left(0\right) + w\left(0\right). \notag
\label{Plant_5}
\end{eqnarray}%

Equation (\ref{Plant_4}) implies that the uncertain term $d\left(q, \dot{q}, u, t\right) $ depends on the system state $\left(q, \dot{q}\right)^T$, control input $u$ and time $t$. Specifically, $a_{1}q\left(t\right) + a_{2}\dot{q}\left(t\right)$ is state-dependent, $b u\left( t\right)$ is input-dependent and $w\left( t\right)$ is time-dependent. The control input coefficient is $1 + b$ with $b$ unknown. In this paper, the control direction is supposed to be time-invariant. 

\textbf{Assumption 1}: $ b \in \left(-1, 1\right) $ to ensure $1 + b >0$.

Use $q_{d}\left( t\right) $, $\dot{q}_{d}$ and $\ddot{q}_{d}$to denote the desired position, the desired velocity and the desired acceleration, respectively. We here assume: (1) the first-order derivative of $w\left( t\right) $ with respect to $t$ exists and is bounded for all $t\geq 0$; (2) the derivatives up to the third-order of the desired position, $\dot{q}_{d}\left( t\right)$, $\ddot{q}_{d}\left( t\right)$ and $ \dddot{q}_{d}\left( t\right)$, are all bounded for $t\geq 0$. Then we define the position error $\tilde{%
	q}\left( t\right)$ and velocity error $\dot{\tilde{q}}\left( t\right) $ as
\begin{eqnarray}
\tilde{q}\left( t\right)  &\triangleq &q_{d}\left(
t\right) -q\left( t\right) ,  \label{PosError} \\
\dot{\tilde{q}}\left( t\right)  &\triangleq & \dot{q}_{d}\left( t\right) -\dot{q}\left( t\right) ,  \label{VelError}
\end{eqnarray}

The PID control (with a desired acceleration compensation term $\ddot{q}_{d}\left( t\right) $) is
\begin{equation}
u\left( t\right) = K_{P}\tilde{q}\left( t\right)  + K_{D}\dot{\tilde{%
		q}}\left( t\right)  + K_{I}\int_{0}^{t}\tilde{q}\left( \tau \right)
d\tau + \ddot{q}_{d}\left( t\right) , t \geq 0, \label{PID_1}
\end{equation}%
where $K_{P}$, $K_{I}$ and $K_{D}$ are the proportional, the integral, and the derivative gains, respectively.

The closed-loop system formed of the controller (\ref{PID_1}) and the plant (\ref
{Plant_3}) with (\ref{Plant_4}) is
\begin{eqnarray}
\ddot{\tilde{q}}\left( t\right) &=&\left[a_{1} - \left(1 + b\right) K_{P}\right]\tilde{q}%
\left(t\right) + \left[a_{2} - \left(1 + b\right) K_{D}\right]\dot{\tilde{q}}\left( t\right) \notag \\
&& - \left(1 + b\right) K_{I}\int_{0}^{t}\tilde{q}\left(\tau \right) d\tau \notag \\
&& - a_{1}q_d\left( t\right) - a_{2}\dot{q}_d \left(t\right) - b \ddot{q}_{d}\left( t\right) - w\left(t\right). \label{ErrorDyn_1}
\end{eqnarray}%

The homogeneous autonomous equation of (\ref{ErrorDyn_1}) is
\begin{eqnarray}
\ddot{\tilde{q}}\left( t\right) &=&\left[a_{1} - \left(1 + b\right) K_{P}\right]\tilde{q}%
\left(t\right) + \left[a_{2} - \left(1 + b\right) K_{D}\right]\dot{\tilde{q}}\left( t\right) \notag \\
&& - \left(1 + b\right) K_{I}\int_{0}^{t}\tilde{q}\left(\tau \right) d\tau. \label{ErrorDyn_2}
\end{eqnarray}%

As for the error trajectories satisfying (\ref{ErrorDyn_1}), we are usually interested in the following two problems.

P1: determining the condition on $K_{P}$, $K_{I}$ and $K_{D}$ under which $\tilde{q}\left(
t\right) $ and $\dot{\tilde{q}}\left( t\right) $ are (globally) bounded for all $t\geq 0$.

P2: tuning $K_{P}$, $K_{I}$ and $K_{D}$ to ensure that $\tilde{q}\left( t\right) $ is ultimately bounded by an arbitrarily small positive number $\varepsilon $, i.e.,%
\begin{equation}
\left\Vert \tilde{q}\left( t\right) \right\Vert \leq
\varepsilon ,\forall t\geq t_{\varepsilon },
\end{equation}%
where $\varepsilon $ denotes an ultimate bound of $\boldsymbol{\tilde{q}}
\left( t\right)$, and $t_{\varepsilon }$ is the corresponding settling
time.

Define the following auxiliary state variable
\begin{equation}
\tilde{q}_{I}\left(t \right) = \int_{0}^{t}\tilde{q}\left(\tau \right) d\tau. \label{Integral-1}
\end{equation}%
Then the integral-differential equation (\ref{ErrorDyn_2}) can be converted into a third-order equation (shown in the following (\ref{ClosedLoop-14})). By applying the well-known Routh-Hurwitz criterion to the characteristic polynomial, we can directly solve the problem P1, and the stability condition can be given as (the detailed analysis process is omitted due to the limitations of space):
\begin{equation}
\left\{
\begin{array}{c}
K_{P} > \frac{a_1}{1 + b}, \\
K_{D} > \frac{a_2}{1 + b}, \\
0 < K_{I} < \left(K_{P} - \frac{a_1}{1 + b}\right)\left[\left(1 + b\right) K_{D} - a_2 \right].
\end{array}%
\right. \label{Condition-1}
\end{equation}%
which is sufficient and necessary to guarantee the exponential stability of the origin of (\ref{ErrorDyn_2}). 

However, the feasible set defined by (\ref{Condition-1}) for gains $K_{P}, K_{I}$, and $K_{D}$ are closely related with the unknown coefficients $a_{1}$, $a_{2}$ and $b$. So, we generally have to  determine $K_{P}, K_{I}$, and $K_{D}$ by the trial and error method. Besides, it is difficult to solve the problem P2 because of the following two facts: 

1) each of the three gains of PID control may affect the ultimate bound of control error, and to the best of our knowledge, this effect cannot be described by a simple relationship, such as a linear or monotone relationship;

2) the unknown uncertainties and exogenous disturbances, i.e. the term $- a_{1}q_d\left( t\right) - a_{2}\dot{q}_d \left(t\right) - b \ddot{q}_{d}\left( t\right) - w\left(t\right)$ in (\ref{ErrorDyn_1}), will also affect the ultimate bound of tracking error. 
 
In the following section, we shall present a simple nonlinear mapping to determine gains $K_{P}, K_{I}$, and $K_{D}$, and then show its applications in solving the problems P1 and P2.

\section{A PID gain mapping: definition, properties and applications}

\subsection{The definition of the gain mapping}
For PID controller (\ref{PID_1}), we define a continuous gain mapping $f: \{\left(k_{p}, k_{d}, T\right) \in R^3 \mid k_{p}>0, k_{d}>0, T>0 \} \mapsto \{\left(K_{P}, K_{I}, K_{D} \right) \in R^3\}$ as follows. 
\begin{equation}
\left\{
\begin{array}{ccc}
K_{P} &=& k_{p}+\frac{k_{d}}{T},\\
K_{D} &=& k_{d}+\frac{1}{T}, \\
K_{I} &=& \frac{k_{p}}{T}.
\end{array}%
\right.  \label{ParaMapping-1}
\end{equation}%
where $k_{p}, k_{d}$ and $T$ are auxiliary parameters. 

Once a set of parameters $\left(k_{p}, k_{d}, T\right)$ are chosen, the idea is to determine the PID gains $K_{P}$, $K_{D}$ and $K_{I}$ by (\ref{ParaMapping-1}) as shown in Fig.1. We shall show that $k_{p}$ and $k_{d}$ can be regarded as the two parameters of a nominal second-order error system and $T$ as the parameter of a first-order UDE. 

\begin{figure}[tbph]
	\centering	
	\includegraphics[width=8cm]{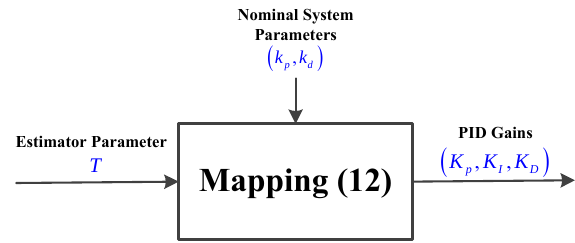}
	\caption{The basic idea to determine PID gains.}
	\label{fig:Idea}
\end{figure}

It is seen from (\ref{ParaMapping-1}) that all the three gains $K_{P}$, $K_{D}$ and $K_{I}$ are associated with parameter $T$.  We have the following computational formulas for the partial derivatives. 
\begin{equation}
\left\{
\begin{array}{ccc}
\frac{\partial K_{P}}{\partial k_{p}} = 1, \frac{\partial K_{P}}{\partial k_{d}} = \frac{1}{T}, \frac{\partial K_{P}}{\partial T} = -\frac{k_{d}}{T^2},\\
\frac{\partial K_{D}}{\partial k_{p}} = 0, \frac{\partial K_{D}}{\partial k_{d}} = 1, \frac{\partial K_{D}}{\partial T} = -\frac{1}{T^2}, \\
\frac{\partial K_{I}}{\partial k_{p}} = \frac{1}{T}, \frac{\partial K_{I}}{\partial k_{d}} = 0, \frac{\partial K_{I}}{\partial T} = -\frac{k_{p}}{T^2}.
\end{array}%
\right.  \label{ParaMapping-2}
\end{equation}%

Under the condition $k_{p}>0$ $k_{d}>0$, and $T>0$, the following three statements are true from (\ref{ParaMapping-1}).

(1) $K_{P}$ depends on $k_{p}$, $k_{d}$ and $T$, and is a linearly increasing function of $k_{p}$ (with fixed $T$ and $\frac{k_{d}}{T}$) or $\frac{1}{T}$ (with fixed $k_{d}$ and $k_{p}$) and $k_{d}$ (with fixed $T$ and $k_{p}$). Thus, the smaller $T$ is or the larger $k_{d}$ is or the larger $k_{p}$, the larger $K_{P}$ is. 

(2) $K_{D}$ is a linearly increasing function of $k_{d}$ (with a fixed $T$) and $\frac{1}{T}$ (with a fixed $k_{d}$). Thus, the smaller $T$ is or the larger $k_{d}$ is, the larger $K_{D}$ is.

(3) $K_{I}$ is a linearly increasing function of $k_{p}$ (with a fixed $T$) and $\frac{1}{T}$ (with a fixed $k_{p}$). Thus, the smaller $T$ is or the larger $k_{p}$ is, the larger $K_{I}$ is.

With (\ref{ParaMapping-1}), the tuning of $K_{P}$, $K_{D}$ and $K_{I}$ can be converted to the tuning of $T$. In particular, \textbf{$K_{P}$, $K_{D}$ and $K_{I}$ determined by (\ref{ParaMapping-1}) will be simultaneously higher if $T$ is adjusted to be smaller}. Thus, when $k_{p}$ and $k_{d}$ are fixed to be certain positive real numbers and $T$ is the single adjustable parameter, we have  

\begin{equation}
\frac{\mathrm{d} K_{P}}{\mathrm{d} T} = -\frac{k_{d}}{T^2}, \frac{\partial K_{D}}{\partial T} = -\frac{1}{T^2},  \frac{\partial K_{I}}{\partial T} = -\frac{k_{p}}{T^2}.
\label{ParaMapping-3}
\end{equation}%

\subsection{The inverse mapping of (\ref{ParaMapping-1})}
Given a set of PID gains $K_{I}$, $K_{D}$, and $K_{P}$, three  questions arise naturally: 

1) do the parameters $k_{p}$, $k_{d}$, $T$ satisfying (\ref{ParaMapping-1}) exist?

2) if existing, how many sets of parameters $\left(k_{p}, k_{d}, T\right)$ exist 

3) if existing, how to compute $k_{p}$, $k_{d}$ and $T$?
     
For the following analysis, we define the cubic polynomial 
\begin{equation}
     p(T) = K_{I}{T}^3 - K_{P}{T^2} + K_{D}T - 1. \label{Polynomial}
\end{equation}
We answer questions 1) and 3) in Lemma 1.

\textbf{Lemma 1}  Given a set of PID gains $\left(K_{P}, K_{I}, K_{D} \right) \in R^3$ with the integral gain $K_{I}>0$, there must exist (at least) one set of parameters $\left(k_{p}, k_{d}, T \right) \in R^3$ with $T>0$ satisfying (\ref{ParaMapping-1}). Furthermore, the set of parameters $\left(k_{p}, k_{d}, T \right) \in R^3$ can be determined by solving the cubic equation $p(T)= 0$ for $T$ and then using the equations $k_{d} = K_{D} - \frac{1}{T}$ and $k_{p} = K_{I}{T}$ to obtain $k_{d}$ and $k_{p}$.

\textbf{Proof}: The second and third equations in (\ref{ParaMapping-1}) are equivalent to
$ k_{d} = K_{D} -\frac{1}{T}$ and $k_{p}= K_{I}{T}$, respectively. Substituting the two equivalent equations into the first equation in (\ref{ParaMapping-1}) yields $K_{P} = K_{I}{T} + \frac{K_{D}T -1}{T^2}$, which is clearly equivalent to the cubic equation $p(T) = 0$. Then, the mapping (\ref{ParaMapping-1}) is equivalent to
\begin{equation}
\left\{
\begin{array}{ccc}
p(T)  &=& 0,\\
k_{d} &=& K_{D} - \frac{1}{T}, \\
k_{p} &=& K_{I}{T},
\end{array}%
\right.  \label{ParaMapping-12}
\end{equation}%
by which we can compute  $\left(k_{p}, k_{d}, T\right)$ from $K_{P}$, $K_{D}$ and $K_{I}$.

The existence and number of the set $\left(k_{p}, k_{d}, T\right)$ satisfying (\ref{ParaMapping-1}) are equivalent to the existence and number of $T$ that fulfills $p(T) = 0$, respectively. The condition $K_{I}> 0$ is sufficient to guarantee the existence of (at least) one positive real $T$ for the cubic equation $p(T) = 0$. This is derived with the continuity of $p(T)$ with respect to $T$ and the fact that $p(0) = - 1 < 0$ and $p(+\infty) > 0$ when $K_{I}> 0$. This end the proof of the lemma. 

To answer the question 2), we apply the following lemma to $p(T) = 0$ by letting $a =K_{I}$, $b = - K_{P}$, $c =K_{D}$, and $d = -1$.

\textbf{Lemma 2} (See Pages 158-159 of \cite{polyanin2006}): The roots of a complete cubic equation
\begin{equation}
f\left(x\right) = ax^3 + bx^2 + cx + d = 0, ~~ a \neq 0,  \label{Cubic_1}
\end{equation}%
are calculated by the formulas
\begin{equation}
x_k = y_k - \frac{b}{3a}, ~~ k= 1, 2, 3 ,  \label{Cubic_2}
\end{equation}%
where $y_k$ are the roots of the incomplete cubic equation
\begin{equation}
f\left(y\right) = y^3 + py + q = 0,  \label{Cubic_3}
\end{equation}%
with the coefficients
\begin{equation}
p = - \frac{1}{3}\left(\frac{b}{a}\right)^2 + \frac{c}{a},  q = \frac{2}{27}\left(\frac{b}{a}\right)^3 - \frac{bc}{3a^2} + \frac{d}{a},  \label{Cubic_4}
\end{equation}%
The number of real roots of the cubic equation depends on the sign of the discriminant $D$ defined as
\begin{equation}
D \triangleq \left(\frac{p}{3}\right)^3 + \left(\frac{q}{2}\right)^2.  \label{Cubic_5}
\end{equation}

Case $D> 0$. There is only one real root.

Case $D<0$. There are three real roots.

Case $D=0$. There is one real root and another real root of double multiplicity (this case is realized for $p = q = 0$).

\subsection{Closed-loop system analysis with mapping (\ref{ParaMapping-1})}
Substituting (\ref{ParaMapping-1}) into (\ref{PID_1}), yields
\begin{eqnarray}
u\left( t\right) &=& \left( k_{p}+\frac{k_{d}}{T}\right) \tilde{q}\left( t\right)
+\left( k_{d}+\frac{1}{T}\right) \dot{\tilde{q}}\left( t\right) \notag \\
 &~~&+\frac{k_{p}}{T%
}\int_{0}^{t}\tilde{q}\left( \tau \right) d\tau +\boldsymbol{%
	\ddot{q}}_{d}\left( t\right), t \geq 0,  \label{PID-3}
\end{eqnarray}

There are totally six elementary components included in (\ref{PID-3}): $k_{p} \tilde{q}$, $\frac{k_{d}}{T}\tilde{q}$, $k_{d}\dot{\tilde{q}}$, $\frac{1}{T}\dot{\tilde{q}}$, $\frac{k_{p}}{T
}\int_{0}^{t}\tilde{q}\left( \tau \right) d\tau$ and $\ddot{\boldsymbol{
		q}}_{d}$. These elementary components can be arbitrarily combined to generate several larger components of $u\left( t\right)$. Here, we rewrite (\ref{PID-3}) into the following simple two-component form
\begin{equation}
u\left( t\right) = u_0\left(t\right) - \hat{d}\left(t\right),  \label{PID-3-1}
\end{equation}
where
\begin{eqnarray}
u_0\left(t\right) &\triangleq& \ddot{\boldsymbol{q}}_{d}\left( t\right)  + k_{p} \tilde{q}\left( t\right)  + k_{d}\dot{\tilde{q}}\left( t\right),  \label{PID-3-2} \\
\hat{d}\left(t\right) &\triangleq& - \frac{k_{d}}{T} \tilde{q}\left(t\right)
- \frac{1}{T} \dot{\tilde{q}}\left( t\right) - \frac{k_{p}}{T%
}\int_{0}^{t}\tilde{q}\left( \tau \right) d\tau.  \label{PID-3-3}
\end{eqnarray}

Clearly, $u_0\left(t\right)$ admits a standard PD control structure with the desired acceleration as feedforward compensation. In the context of this paper, it is called the nominal PD control. Its gains are different from the proportional and derivative gains of the original PID control (\ref{PID_1}).
We will show that $\hat{d}\left(t\right)$ is an estimate of the LUD defined by (\ref{Plant_4}) if $T \to 0$. According to (\ref{PID-3-2}) and (\ref{PID-3-3}), the initial values of $u_0\left(t\right)$ and $\hat{d}\left(t\right) $ are
\begin{eqnarray}
u_0\left(0\right) &=&  \ddot{\boldsymbol{q}}_{d}\left(0\right) + k_{p} \tilde{q}\left(0\right) + k_{d}\dot{\tilde{q}}\left(0\right), \label{PID_3-4-1} \\
\hat{d}\left(0\right) &=& - \frac{k_{d}\tilde{q}\left(0\right) + \dot{\tilde{q}}\left(0\right)}{T}.  \label{PID_3-4-2}
\end{eqnarray}
We here note that $\hat{d}\left(0\right)$ depends on the initial values of tracking error variables as well as the values of parameters $k_d$ and $T$. More importantly, from (\ref{PID_3-4-2}), the smaller $T$ is, the larger $\lvert \hat{d}\left(0\right)\rvert$ is (for the general situations where $k_{d}\tilde{q}\left(0\right) + \dot{\tilde{q}}\left(0\right) \neq 0$), and in particular, $\lvert \hat{d}\left(0\right)\rvert \to \infty$ as $T \to 0$. Since $u\left(0\right) = u_0\left(0\right) - \hat{d}\left(0\right)$ with $u\left(0\right) $ denoting the initial control, a smaller $T$ may results in an initial peaking in $u(t)$.

 Define the LUD estimation error as
\begin{equation}
\tilde{d} \left( t\right) = \hat{d}\left(t\right) - d\left(q,\dot{q}, u, t\right), t \geq 0.   \label{LUDError-1}
\end{equation}%

For uncertain system (\ref{Plant_3}), consider the controller (\ref{PID-3-1}) with (\ref{PID-3-2}) and (\ref{PID-3-3}). The resulting tracking error equation is
\begin{equation}
\ddot{\tilde{q}}\left( t\right) =  - k_{p} \tilde{q}\left( t\right) - k_{d}\dot{\tilde{q}}\left( t\right) + \tilde{d} \left( t\right),
\label{Error-1}
\end{equation}
It is seen that the LUD estimation error $\tilde{d} \left( t\right)$ disturb the trajectories of $\tilde{q}\left( t\right) $. By the definition of $\tilde{d} \left( t\right)$,
\begin{equation}
\dot{\tilde{d}} \left( t\right) = \dot{\hat{d}}\left(t\right) - \dot{d}\left(q,\dot{q}, u, t\right).                                \label{LUDError-2}
\end{equation}%

Consider the derivative of (\ref{PID-3-3}) with the equation (\ref{Error-1}),
\begin{eqnarray}
\dot{\hat{d}}
&=& - \frac{k_{d}}{T} \dot{\tilde{q}}
- \frac{1}{T} \ddot{\tilde{q}} - \frac{k_{p}}{T%
} \tilde{q},  \notag \\
&=&- \frac{k_{d}}{T} \dot{\tilde{q}}
- \frac{1}{T} \left[- k_{p} \tilde{q} - k_{d}\dot{\tilde{q}} + \tilde{d} \right]  - \frac{k_{p}}{T%
} \tilde{q},   \notag \\
&=& - \frac{1}{T} \tilde{d}.                        \label{LUDError-3}
\end{eqnarray}

From the equations (\ref{Plant_4})-(\ref{VelError}) and (\ref{Error-1}),
\begin{eqnarray}
\dot{d}\left(q,\dot{q}, u, t\right) &=& \left(a_{2}k_{p}  - bk_{d}k_{p} \right) \tilde{q}  \notag \\
&~& +\left(a_{2}k_{d} + bk_{p} - a_{1} -  bk_{d}^2  \right) \dot{\tilde{q}} \notag \\
&~&+ \left(bk_{d}+\frac{b}{T} - a_{2} \right) \tilde{d}   \notag \\
&~& + a_{1}\dot{q}_d + a_{2} \ddot{q}_d + b \dddot{q}_{d} + \dot{w},                      \label{LUDError-4}
\end{eqnarray}

Substituting (\ref{LUDError-3}) and (\ref{LUDError-4}) into (\ref{LUDError-2}), gives
\begin{eqnarray}
\dot{\tilde{d}} &=& \left(a_2 - \frac{1+b}{T} - bk_{d} \right) \tilde{d} + \left(bk_{d}k_{p}  - a_{2}k_{p}\right)\tilde{q} \notag \\
&~~&+ \left(a_{1} + bk_{d}^2 - a_{2}k_{d} - bk_{p}\right)\dot{\tilde{q}} \notag \\
&~~&+ a_{1}\dot{q}_d + a_{2} \ddot{q}_d + b \dddot{q}_{d} + \dot{w}. \label{LUDError-5}
\end{eqnarray}

The closed-loop system (\ref{Error-1}) and (\ref{LUDError-5}) can be written into the following standard singular perturbation form
\begin{eqnarray}
\left\{
\begin{array}{ccc}
\dot{e} &=& A_1 e + B_1 \tilde{d},  \\
T\dot{\tilde{d}} &=& A_2 \tilde{d} + T B_2 e + Tu_d,
\end{array}%
\right.  \label{ClosedLoop-12}
\end{eqnarray}%
where $T$ is the perturbation parameter,
\begin{eqnarray}
e\left(t \right) &=& \left[
e_1\left(t \right) ~~ 
e_2\left(t \right)\right]^T \triangleq \left[
\tilde{q}\left(t \right) ~~ 
\dot{\tilde{q}}\left(t \right) \right]^T \in R^2, \\ 
u_d\left(t \right) &=&  a_{1}\dot{q}_d\left(t \right) + a_{2} \ddot{q}_d\left(t \right) + b \dddot{q}_{d}\left(t \right) + \dot{w} \left(t \right)
\end{eqnarray}
and
\begin{eqnarray}
A_1 &=& \left[
\begin{array}{cc}
0 & 1 \\
- k_{p} & - k_{d}
\end{array}%
\right], 
B_1 = \left[
\begin{array}{c}
0 \\
1%
\end{array}%
\right],  \\
A_2 &=& a_2 T - 1 - b - bTk_{d}, \\
B_2 &=&  \left[bk_{d}k_{p}  - a_{2}k_{p}, a_{1} + bk_{d}^2 - a_{2}k_{d} - bk_{p} \right]. \label{ClosedLoop-3}
\end{eqnarray}

Consider the first and second equations in (\ref{ClosedLoop-12}) as the slow and fast subsystems, respectively. Then, by definition, the reduced model for system (\ref{ClosedLoop-12}) is the second-order equation
\begin{equation}
\dot{e}^{*} = A_1 e^{*},~~ e^{*}\left(0\right) = e_0,  \label{ClosedLoop-5}
\end{equation}%
which is exponentially stable provided $k_d > 0$ and $k_p > 0$. 

The boundary layer system, by definition, is
\begin{equation}
\frac{dy}{d\tau} = -\left(1 + b\right) y, ~~y\left(0\right) = \tilde{d}\left(0\right),  \label{BLS_1}
\end{equation}%
where ${\tau} = \frac{t}{T}$. 

The problem P1 is equivalent to derive a condition on $k_p$, $k_d$ and $T$, under which the closed-loop system (\ref{ClosedLoop-12}) is input-to-state stable (ISS) with $e\left(t\right)$ and $\tilde{d}\left(t\right)$ as the state variables, and $u_{d}\left(t\right)$ as the time-varying disturbance input. We here turn to applying an alternative approach -- the singular perturbation approach-- to the closed-loop system (\ref{ClosedLoop-12}). This application will yield direct and simple stability conditions and parameter-turning rules for small steady-state error. The following theorem summarizes our findings. 


\textbf{Theorem 1}: Consider the closed-loop system formed of the
plant (\ref{Plant_3}) and the PID controller (\ref{PID_1}) with the parameter mapping (\ref{ParaMapping-1}).
Under Assumption 1 and the conditions
\begin{equation}
k_p > 0, k_d > 0, \label{Theorem1-1}
\end{equation}%

1) there exists $\bar{T}_1 > 0$ such that for
\begin{equation}
T \in \left(0, \bar{T}_1 \right), \label{Theorem1-2}
\end{equation}
the system (\ref{ClosedLoop-12}) is globally ISS, and all system trajectories are bounded for any initial condition;

2) there is a positive constant $\bar{T}_2 $ such that for all $t \geq 0$, $e_0\in R^2$, $\tilde{d}_0 \in R$, and  $0 <T< \bar{T}_2$, the trajectories of system (\ref{ClosedLoop-12}) satisfy
\begin{eqnarray}
&& e\left(t\right) - e^{*}\left(t\right) = O\left(T\right) \\
&&\tilde{d}\left(t\right) - \hat{y}\left(\frac{t}{T}\right) = O\left(T\right), 
\end{eqnarray}%
where $e^{*}\left(t\right)$ and $\hat{y}\left(\tau\right)$ are the solutions of the reduced and boundary-layer problems (\ref{ClosedLoop-5}) and (\ref{BLS_1}); $e\left(t\right)$ and $\tilde{d}\left(t\right)$ are uniformly ultimately bounded by $b_1 T$ and $b_2T$, respectively, implying that the ultimate bounds of $e\left(t\right)$ and $\tilde{d}\left(t\right)$ tend to zero as $T$ tends to zero, where $b_1$ and $b_2$ are two positive constants.

\textbf{Proof}: By the singular perturbation theory \cite{Khalil2002}, the responses of the two equations in (\ref{ClosedLoop-12}) with a sufficiently small $T$ can exhibits the phenomenon that the state variables $e\left(t \right)$ and $\tilde{d} \left(t \right)$ have different time-scale properties; more specifically, $e\left(t \right)$ and $\tilde{d} \left(t \right)$ are  the slow and fast variables, respectively. We note that the signal $u_d$ included in (\ref{ClosedLoop-12}) depends on the exogenous signals and is bounded by constants independent of $T$. Thus, in order to prove the above statements 1) and 2), we here check the related conditions for singular perturbation systems are satisfied by system (\ref{ClosedLoop-12}). The unforced model corresponding to (\ref{ClosedLoop-12}) is
\begin{eqnarray}
\left\{
\begin{array}{ccc}
\dot{e} &=& A_1 e + B_1 \tilde{d},  \\
T\dot{\tilde{d}} &=& A_2 \tilde{d} + T B_2 e. 
\end{array}%
\right.  \label{ClosedLoop-13}
\end{eqnarray}%
Clearly, the origin is an equilibrium point of (\ref{ClosedLoop-13}). Due to the linearity of (\ref{ClosedLoop-13}), the system (\ref{ClosedLoop-12}) is ISS if and only if the origin is an exponentially stable equilibrium point of (\ref{ClosedLoop-13}). Note the two facts: 1) $A_1$ is Hurwitz under the condition (\ref{Theorem1-1}); 2) the equilibrium point of equation (\ref{BLS_1}), $y = 0$, is globally exponentially stable because $1 + b > 0$ under Assumption 2. Then, by applying Theorem 11.4 in \cite{Khalil2002} to (\ref{ClosedLoop-13}), it is straightforward to prove that there exists $\bar{T}_1 > 0$ such that for all $T <\bar{T}_1$, the origin of (\ref{ClosedLoop-13}) is (globally) exponentially stable and the first statement of the theorem is true.

By the definitions of $e\left(t \right)$,  $\tilde{q}\left(t \right)$ and $\dot{\tilde{q}}\left(t \right)$, the initial value of $e\left(t \right)$ (as the slow variable), denoted by $e_0 \triangleq e\left(0\right)$, is independent of parameter $T$. This property of $e\left(t \right)$ is different from that of $\tilde{d}\left(t\right)$. Setting $T = 0$ in the second equation of (\ref{ClosedLoop-12}), yields the quasi-steady state of variable $\tilde{d}\left(t \right) $: $\tilde{d}\left(t \right) = 0, ~~ t \geq 0$. Since $1 + b>0$ by Assumption 1,  the origin of the boundary-layer model is globally exponentially stable uniformly in $(t, e)$. Up to now, all the conditions required by Theorem 11.2 in \cite{Khalil2002} have been verified. By applying the results of the theorem to the above singularly perturbed equation, we readily conclude: for any $1 + b>0$, the second equation in (\ref{ClosedLoop-12}) has a unique solution $\tilde{d}(t, T)$ and so will the first equation in (\ref{ClosedLoop-12}); and the statement 2) is true for any  sufficiently small $T>0$. This is the end of Theorem 1. 

The following result follows directly from Theorem 1.

\textbf{Lemma 3}: Consider the closed-loop system formed of the
plant (\ref{Plant_3}) and the PID controller (\ref{PID_1}) with the parameter mapping (\ref{ParaMapping-1}) under Assumption 1 and the parameter conditions $k_p > 0, k_d > 0$. If $ \dot{q}_d\left(t \right) \to 0 $
and $ \dot{w}\left(t \right) \to 0 $ as $ t \to +\infty $, then $ \dot{u}_d \left(t \right) \to 0 $, and $e\left(t \right)$ and $\tilde{q} \left(t \right)$ converge to zero exponentially (i.e.,
the origin of system (\ref{ClosedLoop-12}) is globally exponentially stable), provided $ T>0 $ is sufficiently small.

\subsection{Comparing the analysis and design procedure}
The closed-loop third-order equation equivalent to (\ref{ErrorDyn_1}) is given in terms of the state variables $e\in R^2$ and $\tilde{q}_{I}\in R $ as
\begin{eqnarray}
\dot{\tilde{q}}_I &=& {e}_1, \notag \\
\dot{e}_1 &=& {e}_2,  \notag \\
\dot{e}_2 &=& -\left(1 + b\right) K_{I}\tilde{q}_{I} + \left[a_{1} - \left(1 + b\right) K_{P}\right]e_{1}   \notag \\
 &~& + \left[a_{2} - \left(1 + b\right) K_{D}\right]e_{2} + u,  \label{ClosedLoop-14}
\end{eqnarray}%
with 
\begin{equation}
u \triangleq - a_{1}q_d\left( t\right) - a_{2}\dot{q}_d \left(t\right) - b \ddot{q}_{d}\left( t\right) - w\left(t\right). \label{ClosedLoop-15}
\end{equation}%
as the disturbance input, and the corresponding system matrix given in a canonical form as
\begin{align}
A&=\left[
\begin{array}{ccc}
0 & 1 & 0 \\
0 & 0 & 1 \\
-\left(1 + b\right) K_{I} & a_{1} - \left(1 + b\right) K_{P} & a_{2} - \left(1 + b\right) K_{D}%
\end{array}%
\right],  \notag  \\
B&=\left[0~~~~ 0~~~~ 1 \right]^T.   \label{A_2}
\end{align}

Without using the mapping (\ref{ParaMapping-1}), the system stability condition are given in terms of the inequalities  (\ref{Condition-1}). Since the unknown coefficients $a_1$, $a_2$ and $b$ are involved, these inequalities are not of much use for tuning the PID gains to guarantee the system stability. In contrast, by applying the mapping (\ref{ParaMapping-1}), the stability condition can be simply reduced to (as indicated by Theorem 1)
\begin{equation}
k_p > 0, k_d >0, 0<T<\bar{T}_1.   \label{StabilityCondition-1}
\end{equation} 
The PID gain tuning is converted to the determination of $\bar{T}_1$. 

By applying Lemma 4 present in Appendix to the third-order equation, we can readily obtain an ultimate bound of $\tilde{q} \left( t\right) $ as given by the following inequality
\begin{eqnarray}
\left\Vert \tilde{q} \left( t\right) \right\Vert _{2} &\leq & \frac{2 \left\Vert
	B\right\Vert_{2}\left\Vert u\right\Vert_{\infty}} {\theta} \sqrt{\frac{%
		(\lambda_{\max }\left( P\right))^3}{\lambda _{\min}\left( P\right)}},
\forall t\geq t_s  \label{Ultimate_Bound_1}
\end{eqnarray}%
where $0<\theta <1$ and $P$ is a symmetric positive-definite matrix
satisfying the Lyapunov equation%
\begin{equation}
PA+A^{T}P=-I_{n}.
\end{equation}

From (\ref{Ultimate_Bound_1}) we see that the ultimate bound of $\tilde{q} \left( t\right) $ is a linear function of $\sqrt{\frac{(\lambda_{\max }\left( P\right))^3}{\lambda _{\min}\left( P\right)}}$. Thus, we can reduce the stead-state error by properly choosing $K_{P}$, $K_{D}$ and $K_{I}$ to render  $\sqrt{\frac{(\lambda_{\max }\left( P\right))^3}{\lambda _{\min}\left( P\right)}}$ small. This idea is intuitive but not feasible because of two facts: 1) the relationships between PID gains ($K_{P}$, $K_{D}$ and $K_{I}$) and $\sqrt{\frac{(\lambda_{\max }\left( P\right))^3}{\lambda _{\min}\left( P\right)}}$ is nonlinear and not explicit; 2) the unknown coefficients $a_1$, $a_2$ and $b$ are involved in the entries of system matrix $A$ and affect the value of $\sqrt{\frac{(\lambda_{\max }\left( P\right))^3}{\lambda _{\min}\left( P\right)}}$. In contrast, as shown in Theorem 1, the steady-state error can be effectively reduced by choosing a smaller $T$. The idea of regulating the steady-state performance by tuning $T$ is simple, intuitive and feasible. 

By comparing (\ref{ClosedLoop-12})  and (\ref{ClosedLoop-14}), we conclude that with the mapping (\ref{ParaMapping-1}), the original third-order error system is converted to a cascade connection of two lower-order subsystems. The subsystem describing the dynamics of the uncertainty estimation error $\tilde{d}\left( t\right)$ is fast and first-order, whereas the subsystem describing the dynamics of the tracking error trajectory $\tilde{e}\left( t\right)$ is slow and second-order. This idea for third-order equation decomposition is shown in Fig. 2.

\begin{figure}[tbph]
	\centering	
	\includegraphics[width=7cm]{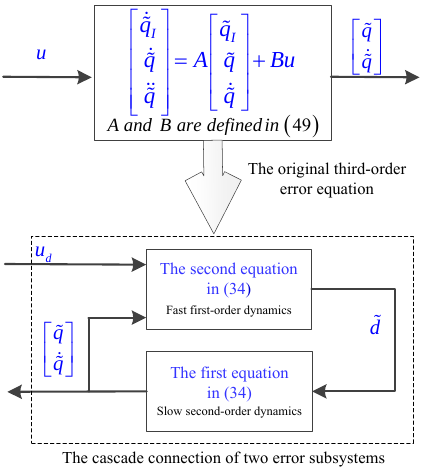}
	\caption{The decomposition result by the mapping (\ref{ParaMapping-1}).}
	\label{fig:Idea}
\end{figure}

By viewing (\ref{PID-3-2}) as a nominal control and (\ref{PID-3-3}) as a UDE, the PID gain tuning procedure is implemented in three steps:

Step 1. Determine the auxiliary parameters $k_d$ and $k_p$ for the nominal PD controller (\ref{PID-3-2}).

Step 2. Choose proper auxiliary parameter  $T$ for the UDE (\ref{PID-3-3}) to recover the performance of the nominal error system.

Step 3. Determine the set of PID gains $\left(K_{P}, K_{I}, K_{D} \right)$ by the relationship (\ref{ParaMapping-1}).

\section{Simulation and experimental results}

The gain mapping is applied to tune the gains of the controller implemented on a 3-DOF helicopter platform \cite{Quanser2006}. The whole controller includes the feedback linearization part and the PID controller part. As shown in \cite{zhu2015CEP} and \cite{MIT2004}, the platform suffers from severe input disturbances (due to the equipped active disturbing system) and the uncertainties in damping coefficient. The nominal parameters of the platform for controller design are reported in \cite{zhu2015CEP}. 

In simulation and experiments, the desired attitude trajectories are given (in degree) as
\begin{eqnarray}
\alpha _{d}\left( t\right) &=&5.73(\sin(0.25t)+\sin(0.5t) - 0.33),  \notag \\
\beta _{d}\left( t\right) &=&15\sin (0.63t),   \label{DesiredTraj}
\end{eqnarray}%
and the initial states of the platform are
\begin{eqnarray}
\left( \alpha(0),\dot{\alpha}(0),\beta\left( 0\right) ,\dot{\beta}\left( 0\right) \right) &=&\left(-25.7,0,0,0\right). \label{Inistial_States}
\end{eqnarray}

\subsection{Simulation results}

For comparison purpose, we consider three sets of PID gains, and two kinds of disturbances: constant disturbance (d1) and non-constant disturbance (d2). For the former disturbance, $d_{\alpha}=0.345$ and $d_{\beta} = 0.015$. For the latter, $d_{\alpha}=0.345\cos(t)$ and $d_{\beta}= 0.015\cos(t)$. The performance of each PID control against the two kinds of disturbances is examined. Due to space limitations, only the results in elevation channel are shown. The PID control gains for elevation channel and the ultimate bounds (UBs) of the tracking errors are compared in TATLE I. In addition, the error responses are detailed in Figs. 3 and 4. 

We see from Fig.3 that each PID control achieves zero-error tracking under the constant disturbance. This result is consistent Lemma 3. For the cases with non-constant disturbances, by comparing the results of P1 and P2 shown in Fig.4, we observe that a larger integral gain $K_I$ results in a larger ultimate bound of error, whereas by comparing the results of P2 and P3, a smaller $K_I$ results in a larger ultimate bound of error. Thus, choosing a larger integral gain alone is not always effective in reducing the tracking error. The difference in the ultimate bounds of P2 and P3 (sharing the parameter $T=0.4$) shows the effect of $(k_p, k_d)$ on steady-state errors. Noting that P1 and P3 have the same PD gain $(k_p, k_d) = (1, 2)$ and a different UDE parameter $T$, we conclude that a smaller $T$ indeed leads to a smaller ultimate bound of the error trajectories.

\begin{table}[tph]
	\caption{PID gains and the simulation results}
	\label{tb:Para}
	\centering
	\begin{tabular}{cccccc}
		\hline\hline
		No. &$(K_P, K_I, K_D)$& $(k_p, k_d, T)$ & UBs(d1) & UBs(d2) \\ \hline
		P1 & (21, 10, 12) & (1, 2, 0.1) & 0 & 0.95 deg \\
		P2 & (16, 15, 6.5) & (6, 4, 0.4) & 0 & 1.11 deg \\
		P3 & (6, 2.5, 4.5) & (1, 2, 0.4) & 0 & 3.55 deg\\
		\hline\hline
	\end{tabular}%
\end{table}

\subsection{Experimental results}
For comparison purpose, we here implement two PID controllers on the platform, with parameters P4 and P5 shown in TABLE II. The gains of both PID controllers are computed using the mapping (\ref{ParaMapping-1}) with the same PD gain $(k_p, k_d) = (2, 1.5)$ and different UDE gains $T=0.5$ and $T=0.1$, respectively. The ultimate bounds of tracking error are compared in TABLE II, the detailed error responses are shown in Fig. 5, and the resulting input voltages are shown in Figs. 6 and 7.
\begin{table}[tph]
	\caption{PID gains and the experimental results}
	\label{tb:Para}
	\centering
	\begin{tabular}{ccccc}
		\hline\hline
		No. &$(K_P, K_I, K_D)$& $(k_p, k_d, T)$ & UBs(d3) \\ \hline
		P4 & (5, 4, 3.5) & (2, 1.5, 0.5) & 1.78 deg \\
		P5 & (17, 20, 11.5) & (2, 1.5, 0.1) & 0.46 deg \\
		\hline\hline
	\end{tabular}%
\end{table}

\begin{figure}[tbph]
	\centering	
	\includegraphics[scale = 0.3]{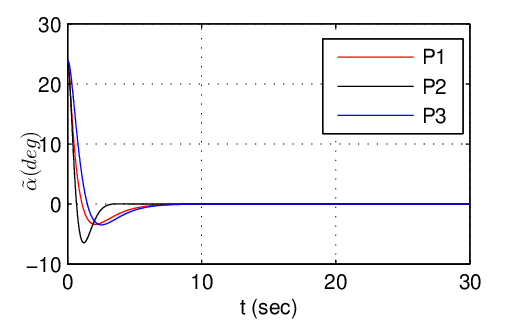}
	\caption{Error trajectories under constant disturbances}
	\label{fig:Sim2}
\end{figure}

\begin{figure}[tbph]
	\centering	
	\includegraphics[scale = 0.3]{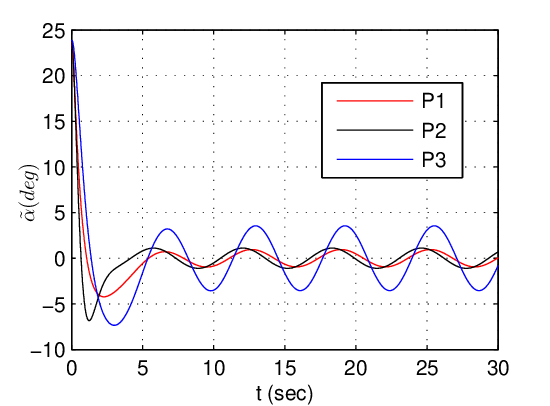}
	\caption{Error trajectories under non-constant disturbances}
	\label{fig:Sim3}
\end{figure}

\begin{figure}[tbph]
	\centering	
	\includegraphics[scale = 0.3]{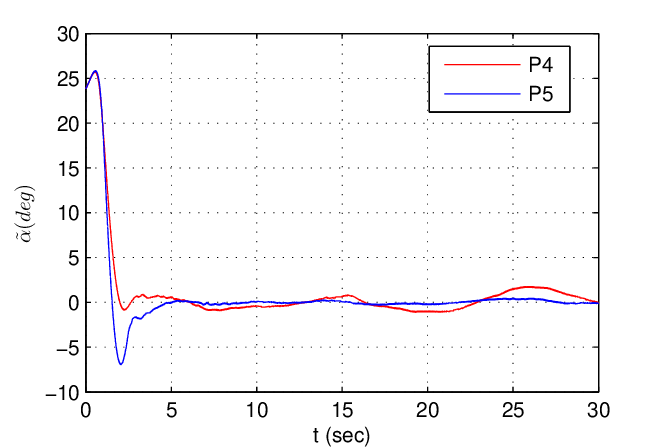}
	\caption{Error trajectories obtained in experiments}
	\label{fig:Exp1}
\end{figure}

\begin{figure}[tbph]
	\centering	
	\includegraphics[scale = 0.3]{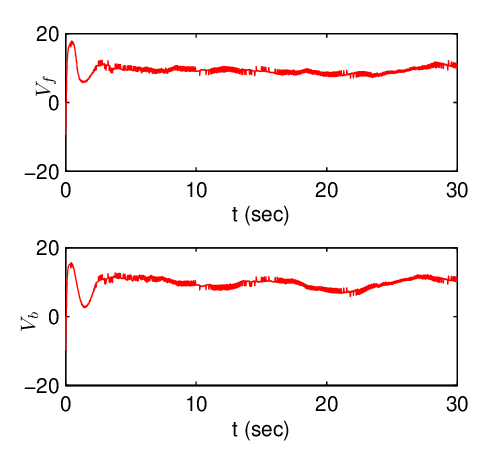}
	\caption{Input voltages resulting from P4}
	\label{fig:Exp2}
\end{figure}

\begin{figure}[tbph]
	\centering	
	\includegraphics[scale = 0.3]{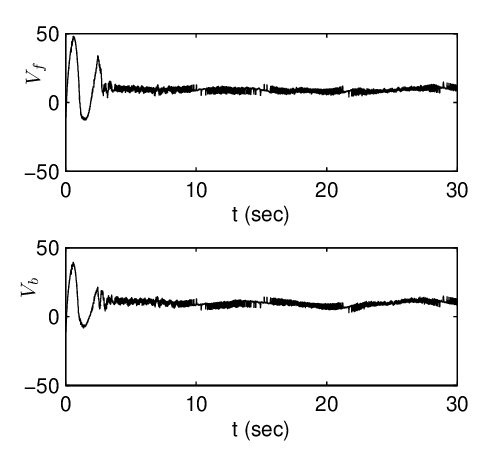}
	\caption{Input voltages resulting from P5}
	\label{fig:Exp3}
\end{figure}

It is seen that the effects of actual disturbances are attenuated by the two PID controllers in different degree. Specifically, a much smaller ultimate bound of the tracking error is obtained with a smaller $T$. Noting that a smaller $T$ makes the PID gains higher, we conclude from Figs. 6 and 7 that the use of a smaller $T$ leads to input voltages of higher transient magnitude. This in turn implies the control cost resulting from the use of higher-gain PID controllers.

\section{CONCLUSIONS}

In this paper,, we have introduced a novel gain mapping for PID control. With the mapping, the PID control is shown to be equivalent to a new PD control serving as a nominal control, plus an uncertainty and disturbance compensator used to recover the nominal performance. We have proven that both closed-loop stability and steady-state error performance are explicitly related to an auxiliary parameter. Then PID control can be understood, designed, and tuned in a 2-DoF control framework. We believe that the robustness of the PID controller with respect to model uncertainties and exogenous disturbance can be clearly understood. More importantly, the three-parameter tuning problem of the PID controller can be reduced to a single-parameter tuning problem by applying the mapping. Future work includes the application of the mapping to general non-linear systems.



\section*{APPENDIX}

The following result is applied to show the ultimate bounds of the solutions of linear state equations.

\textbf{Lemma 4} (Lemma 1 in \cite{zhu2015CEP}): Consider the state solution
$x\left( t\right) $ of the linear time-invariant equation%
\begin{equation}
\dot{x}=Ax+Bu,  \label{Sys_Lemma1}
\end{equation}%
where $x\in R^{n}$ is the state, $u\in R$ is the continuously differentiable
input and matrices $A$ and $B$ have compatible dimensions. If $A$ is
Hurwitz, then

1) system (\ref{Sys_Lemma1}) is globally input-to-state stable (GISS), that
is, if $u$ is bounded for all $t$ (i.e., $u\in \mathcal{L}_{\infty }$), then
$x\left( t\right) $ with any initial state $x\left( t_{0}\right) $ is also
bounded for all $t$;

2) there exist a class $\mathcal{KI}$ function $\mathcal{\beta }$ and a time
$t_s \geq t_{0}$ such that for any $x\left(t_{0}\right) $,
\begin{eqnarray}
\left\Vert x\left( t\right) \right\Vert _{2} &\leq &\mathcal{\beta }\left(
\left\Vert x\left( 0\right) \right\Vert _{2},t\right), \forall t_{0}\leq
t\leq t_s  \label{Bound_1} \\
\left\Vert x\left( t\right) \right\Vert _{2} &\leq & \frac{2 \left\Vert
	B\right\Vert_{2}\left\Vert u\right\Vert_{\infty}} {\theta} \sqrt{\frac{%
		(\lambda_{\max }\left( P\right))^3}{\lambda _{\min}\left( P\right)}},
\forall t\geq t_s  \label{Appendix_1}
\end{eqnarray}%
where $0<\theta <1$ and $P$ is a symmetric positive-definite matrix
satisfying the Lyapunov equation%
\begin{equation}
PA+A^{T}P=-I_{n}.
\end{equation}

\section*{ACKNOWLEDGMENT}
The authors would like to thank Prof. Xiang Chen at Department of Electrical and Computer Engineering, University of Windsor for constructive discussions on the paper.

\bibliographystyle{IEEEtran}
\balance
\bibliography{ref}

\end{document}